\documentclass[preprint, aps]{revtex4}
\usepackage{graphicx}

\newcommand{\ba}{\begin{eqnarray}}
\newcommand{\ea}{\end{eqnarray}}

\begin{document}

\title{Evolution of a black hole at the center of GRB}

\author{Hyun Kyu \surname{Lee} $^{a,b}$\footnote{e-mail :
hklee@hepth.hanyang.ac.kr} and  Hui-Kyung \surname{Kim} $^{a}$
\footnote{e-mail : black072@hepth.hanyang.ac.kr}}
\affiliation{ $^a$ Department of Physics, Hanyang University, Seoul 133-791, Korea \\
and \\ $^b$ Asia Pacific Center for Theoretical Physics, Pohang Kyungbuk 790-784, Korea}

\begin{abstract}

Using a simplified model of a black hole-accretion disk system which
is dominated by Poynting flux, the evolution of the central black
hole which is supposed to be powering GRB is discussed.  It is
demonstrated explicitly that there is a lower limit on the angular
momentum parameter for a given GRB energy.  It is found that the
most energetic GRBs can only accommodate relatively rapid-rotating
black holes at the center.   For a set of GRBs for which  the
isotropic energies and $T_{90}$s are known,   the effect of the disk mass and the magnetic field on the horizon are
discussed quantitatively. It is found that the magnetic field has
little influence on the energy but affects the GRB duration as
expected. The role of the disk mass is found to be significant in
determining  both the energy and the duration.

\vskip .3cm

\noindent PACS numbers: 97.10.Gz, 97.60.Lf

\end{abstract}

\maketitle

\section{Introduction}
 The discovery and the recent observations followed afterwards  indicate
 that the central region  powering gamma ray burst is rather compact in size,
less than $10^8$ cm,  from the studies of energies and temporal
structures\cite{piran}.    Merging binary compact objects,
hypernovae and rotating black holes have been considered to be
among the viable candidates for a GRB central engine.  The common
feature of these models is the formation or the existence of a
black hole at the central region.  However, the gamma ray bursts
or the afterglows followed do not  provide any direct
observational informations on the central region and we do not
know the exact nature of  the black hole at the center.  Hence
although it is very important and interesting to investigate the
central object itself,  we should rely on the indirect method of
using a particular model, in which the physical properties  of
black holes can be inferred from the observational data.

Among the models proposed so far, we choose a model in which the
rotational energy of the black hole is responsible for powering
GRB\cite{lwb}. The mechanism of tapping rotational energy  from
the black hole has been known as Blandford-Znajek
mechanism\cite{bz}\cite{TPM}, in which the rotational energy is
extracted out to the loading region via the magnetic flux which
threads on the horizon.  It is easy to demonstrate that the strong
magnetic field of $\sim10^{15}$gauss is consistent with the
essential features of GRBs\cite{lwb}.

It is well known that the  black hole itself cannot keep the
magnetic field on the horizon. It disappears very rapidly if not
the environment keeps them from disappearing.  Recently, it is
shown that the magnetic flux on the strongly magnetized object can
be maintained during the collapsing process together with the
electric charge onto the black hole\cite{llp}. When the
gravitationally unstable  object collapses into black hole the
most natural environment is the accretion disk/torus which emerges
together with the central black hole and it can provide  the
magnetoshpere which keeps the magnetic flux from disappearing.
Because of the accretion the energy and the angular
momentum  are carried into the black hole while the Poynting flux
carries away energy and angular momentum out of the black
hole\cite{lbw}, the evolution of a black hole depends not only on
the Poynting flux but also on the accretion.

In this work we make an attempt to infer the evolution of the
central black hole   during the gamma rays  are bursting using a
simple model of a black hole-accretion disk system suggested in
\cite{lk}, in  which the accretion is dominated by the Poynting
flux\cite{blandford}\cite{hkl}. In section II, the simple minded model is
sketched with emphasis on the parameters which govern the
evolution of the black hole. We take the initial values of the
magnetic field on the horizon and the disk mass as parameters
while the black hole mass is initially taken to be a typical mass,
$7M_{\odot}$, suggested from the  black hole binary  systems. The
evolution of the angular momentum parameter $\tilde{a}$ is
discussed in detail in section III. With finite mass of accretion
disk, $M_D \leq 7M_{\odot}$, the initially rapid-rotating black
hole is expected to remain rotating even at the end
of GRB. But the initially slow-rotating black hole is found to
eventually stop rotating with GRB.  In section IV, the total
energy out of the rotating black hole is calculated and compared
to  GRBs for which the isotropic energies are known\cite{grb}.
The lower limits of the angular momentum parameter of the black holes for the
corresponding GRBs are calculated. For the most energetic
GRBs, $E_{iso}
> .1 M_{\odot}$, the lower limit is found to be rather high,
$\tilde{a}(0) >0.3$.  In section V, the evolution of the black hole
is discussed for GRBs with known $T_{90}$.  Identifying
$E_{iso}$ as 90\% of energy extraction and  $T_{90}$ as the time
taken to extract out $E_{iso}$, the corresponding sets of $B_H(0)$
and $M_D(0)$ are determined to discuss the evolution of the
corresponding black holes.  The discussions are given in section
VI.

\section{Brief sketch of a model}

In this work, the energy of GRB  is supposed  to be powered by
Poynting flux out of  the rotating black hole surrounded by
the magnetized accretion disk\cite{lwb}, Fig. 1.  To see the effects of the
magnetic field on the evolution of the black-hole-accretion disk,
 it is assumed that the  accretion flow is driven mainly by
the strong magnetic field which carries angular momentum by
Poynting flux\cite{lk}\cite{blandford}.  Then we can have an
analytic formulation of the black hole evolution.  The essential
feature of the  evolution model discussed in  ref \cite{lk} is
briefly reviewed in the following.

  The evolution rates  of  the black hole mass ($\dot{M})$ and the angular
momentum ($\dot{J}$) are determined both  by the energy and the
angular momentum  accreted, which increase the mass and the
angular momentum, and also by the Blandford-Znajek power in the
opposite direction. From  the energy and angular momentum
conservation, we get the rates of change for the mass and the
angular momentum given by
\ba
\dot{M} &=&  - P_{BZ} + \dot{M}_{+} \tilde{E}, \\ \dot{J} &=& -
\frac{P_{BZ}}{ \Omega_{F}}+ \dot{M}_{+} \tilde{l} \ea where
$\dot{M}_{+}$ is the mass accretion rate and the specific energy
and angular momentum of the accreting matter are denoted as
$\tilde{E}$ and  $\tilde{l}$ respectively. Since $P_{BZ}$ is
proportional to  $\dot{M}_{+}$ in this particular model\cite{lk}, one can
get
\ba
P_{BZ}  =  ( \tilde{E}- \tilde{P}_E) \dot{M}_{+},
\label{mbdot} \,\,\, \dot{J} = M \dot{M}_{+}({\cal Z}_{l}-
\tilde{P}_l), \ea
where
\ba
\tilde{E} = \frac{Z^{2}-2Z+ \tilde{a} \sqrt{Z}}{Z \sqrt{Z^{2}-3Z+2 \tilde{a} \sqrt{Z}}}, \,\,\,
\tilde{P}_E = \frac{f(h)H^2}{4Z} [1 + (\tilde{a}/Z)^2(1+ 2/Z)] \label{ptilde},\\
\tilde{l}=M \tilde{\cal Z}_{l}(\tilde{a}), \,\,\,
\tilde{\cal Z}_{l}(\tilde{a}) = \frac{Z^{2}-2 \tilde{a} \sqrt{Z}
+\tilde{a}^{2}}{\sqrt{Z(Z^2-3Z+2 \tilde{a} \sqrt{z})}}, \\
 \tilde{P}_l = \frac{f(h)H^3}{\tilde{a}Z} [1 + (\tilde{a}/Z)^2(1+ 2/Z)], \label{ptildl}
\ea
and
\ba
f(h)= \frac{1 + h^{2}}{h^{2}}[(h + \frac{1}{h}) \arctan{h} - 1], \,\,\, h = \frac{ \tilde{a}}{H}, \,\,\,  H=1 + \sqrt{1 - \tilde{a}^2}
\ea
It is assumed that the
inner edge of the accretion disk is the last stable
orbit\cite{shapiro} defined  by
\ba
r_{in} =  \,  Z M ,
\ea

Since the presence of an accretion disk with an appreciable magnetic
 field is essential for the magnetic field on the horizon, the evolution
of the accretion disk is also responsible for the evolution of the magnetic field on the
disk and $B_H$.

For the numerical calculations in this work, we assume that the effects of mass loss from the disk can be incorporated into the the
time dependence of the magnetic field in the following form:
\ba
B_H^2 = B_H^2(0)D(t), \ea where
\ba
 D(t) = 1 - (\int^t_0 \dot{M}_+)/M_{D}(0).
\ea We take $D(t)$ to be vanishing as the total accreting mass
becomes the initial disk mass, which is supposed be  typically of
solar  mass for an accretion disk emerging out of binary - merging
processes.

\section{Evolution of angular momentum}

  The  rotation of a black hole can be described either by angular
  momentum
  $J$, or specific angular momentum $a(=J/M)$ or angular momentum
  parameter $\tilde{a}(=a/M)$. In this work we will take the angular momentum
  parameter $\tilde{a}$ which is a dimensionless quantity to
  represent the rotation of the black
hole.

The evolution of the angular momentum parameter $\tilde{a}$ can be
obtained from the angular momentum evolution as
\ba
 \frac{ \dot{J}}{J} &=& \frac{ \dot{a}}{a}+ \frac{ \dot{M}}{M}
\ea  Then, using Eq. (\ref{mbdot}), we get
\ba
\dot{ \tilde{a}} = A \frac{\dot{M}_+}{ M}\label{sign} \ea  where
the sign factor $A$ defined by
  \ba
   A=({\cal
Z}_{l}-\tilde{P}_l) -2\tilde{a}(\tilde{E} - \tilde{P}_E) \ea
determines whether $\tilde{a}$ is decreasing or increasing. As
shown in Fig. 2, the sign factor is  negative  for the entire
range of $\tilde{a}$. It means that
     the rate of change of the angular momentum
  parameter $\dot{\tilde{a}}$, is always negative in
  this model described in the previous section.
  It should be noted that this sign factor  is independent of
  the parameters of the model. It is the intrinsic
  feature of the model.

  The sign factor has a maximum at
  $\tilde{a} \rightarrow  \tilde{a}_m =  .57$
  but still less than zero.  It is analogous
  to the energy barrier to be overcome in particle dynamics.  If the evolution
  starts with
  $\tilde{a} > .57$  one should wait long time until  the black
  hole stops rotating. It is because  when it is crossing the maximum value,
   $\tilde{a}_m$, the
  process becomes  very slow.  With the
  finite mass of the accretion disk, the evolution ends  when the
  disk disappears completely into the black hole. In this case however
  the evolution may end up with the rotating black hole as one can see
  in Fig. 3.  On the other hand, if the evolution starts with
  $\tilde{a} < .57$  the slow-down process accelerates in the beginning and the
  black hole eventually stops rotating in a much shorter time.
  However
  it depends also on the evolution of the disk.  Since the life
  time of the disk depends  essentially on the initial  mass,
  we can define,  for a given mass of accretion disk,
  $\tilde{a}_{c}$(Fig. 4): a black hole start with $\tilde{a} <
  \tilde{a}_{c}$ ends up with non-rotating black hole.

\section{Energy of a  gamma ray burst}

In this work, the energy of GRB is supposed to be powered by the
Poynting flux out of the rotating black hole at the center of a
black hole - accretion disk system.   Essentially the energy  is a
part of the  rotational energy of the black hole.  The rotational
energy of the disk is accreted into the black hole and then it is
processed to be  a part of the black hole's rotational energy.
Therefore the source of the energy delivered to GRB is not only
the black hole's initial rotational energy but also the rotational energy
of the disk.

 The duration of the GRB is identified to be the duration of Poynting
flux from the system.  The Blandford-Znajek process becomes
ineffective  either when there is no magnetic field on the black
hole or it does not rotate.   If the system starts with the larger
disk mass, the life time is longer for a given initial magnetic
field.  Therefore the energy carried out until the
Blandford-Znajek process stops is expected to be larger for the
larger disk mass system. Numerical calculations in Fig. 5 shows
that the energy is increasing with the initial disk mass,
$M_D(0)$.

However for $\tilde{a} < \tilde{a}_{c}$, the energy reaches maximum
at a certain value of disk mass,  $M_{D}^{c}$, and shows little change
for the larger disk mass.  It is because the rotation of the black
hole slows sufficiently down for the  Blandford-Znajek process to
be   ineffective as discussed in the previous section.   When the
system starts with $M_{D}(0) > M_{D}^{c}$,  the black hole is slow down
substantially before the life time of the accretion disk.  And the
GRB duration is determined by the time when black hole stops
rotating not by the life time of the disk.  The radial current
in the disk vanishes and there is no magnetic breaking hence no
accretion onto the black hole.   For $\tilde{a}
> \tilde{a}_{c}$, the disk disappears into the black hole while the
remaining black hole is still rotating. But there is no Poynting
flux because the magnetic field supported by the disk disappears
altogether. Then the duration of GRB is determined by the disk
life time.

Since we are interested in the system of black hole - accretion
disk which emerges during a collapse or merging process of stellar
objects, $M_D$ might not be much greater than the black hole mass
in the system. And for numerical calculation we take the upper
limit to be $7M_{\odot}$ in this work.  It gives upper limits of
the energy out of the system as shown in Fig. 6 for given initial
values of the angular momentum  parameter.

Compared to the isotropic energies of GRB, one can
estimate the lower limits of the initial angular momentum
parameters, $\tilde{a}_{l}$, of the black holes at the center of the GRB in this
model, Table I.  One can see that the energetic GRBs
require rapidly rotating black holes at the
center: for example  $\tilde{a}(0) \ge 0.45$ for GRB 990123.

\section{Evolution of a black hole}

The effect of the magnetic field on the total energy out of the
system is not significant as shown in Fig. 7.
But the strength of the magnetic field is directly involved in
determining the rate of energy carried out along the magnetic
flux.  The stronger magnetic field extracts  energy more rapidly
than the weaker magnetic field.  Hence  the detailed variation of
the GRB duration is expected to be related to the magnetic field
structure of the system. In this work it is parameterized by the
initial value on the horizon, $B_{H}(0)$.

The determination of the  duration of GRB  from the observational
data is relatively complicated than the isotropic energy.  In this
work,  as a first trial, we take $T_{90}$ as the duration  for 90
$\%$ of the total Poynting flux energy to be  carried out. Among
the GRBs with inferred isotropic energies in Table I,  we choose
six of them in Table II for which $T_{90}$s are well defined.

The evolution of a black hole is determined by the mass and
angular momentum.  For  the system  with $\tilde{a}(0) =.8$,
it is found that the
final mass of the black holes are found to be increasing: $ \Delta
M_{BH} = .23 M_{\odot}$ upto $2.8M_{\odot}$. Since $\tilde{a}(0) =.8
> \tilde{a}_m$, the final black holes are expected  to be
rotating even after GRB with smaller angular momentum parameters
than the initial values. For $\tilde{a}(0) =.3 $, which is smaller
than the lower limits of GRB 990123, 990506, 991216, 000131
which require more rapidly
rotating black holes at the center,  the less energetic GRBs,
990510 and 991208, can
accommodate the slowly rotating black holes. In this case,
however, the appropriate disk masses are found to be smaller than
$M_D^c$ and the black holes remain rotating after GRB. It is
because $\tilde{a}(0)$ is greater than $\tilde{a}_c$ for
the corresponding disk mass.

\section{Discussion}

Using a schematic model for a central engine of GRB, which
consists of a black hole and an accretion disk\cite{lk} at the
center, the evolution of the central black hole is discussed.  The
accretion is assumed to be dominated by the Poynting flux out of
the disk and the GRB is supposed to be powered by the Poynting
flux which extracts out a part of the rotational
energy of a black hole.
      It is found that the evolution of the rotation
parameterized by $\tilde{a}(t)$ shows different patterns depending
 on  the initial value, $\tilde{a}(0)$.   For a given disk mass,
the black hole with  $\tilde{a}(0) < .57$  much more rapidly
approaches to non-rotating black hole than with  $\tilde{a}(0) >
.57$.   It is also shown that there is a maximum energy for GRB
for a given $\tilde{a}(0)$. Hence one can infer the lower limit of
the angular momentum parameter for the central black hole.  The
most energetic GRBs, are found to be able to
accommodate only rapidly rotating black holes , $\tilde{a}(0)> 0.4$.

     The effect of the magnetic field on the total energy for GRB  is
found to be not significant compared to the disk mass.  However
since the stronger magnetic field extract  energy more rapidly
than the weaker magnetic field, the detailed variation of the GRB
duration is found to be due to  the magnetic field structure of
the system. The role of the disk mass in this model is significant
both in determining the energy and the duration of GRB.  Within the
range of the parameters  used to fit a set of GRBs for which
isotropic energy and $T_{90}$ are well determined, the final black
holes are found to become  more massive  than the initial values but
 with  smaller angular momentum parameters.   This observation is
 consistent with the  general feature expected for a system of  a black hole
- accretion disk with finite size and life time.
As it is mentioned the analysis is limited to GRBs with $T_{90}$
determined in this work.
But provided with a systematic way of determining
GRB duration time from the observational data,
statistically the more meaningful conclusion can be made on the evolution of
the black hole at the center of GRB and it remains as a future work.
 The characteristic
 feature of this model is the sign factor $A$ determined by the
 magnetically dominated accretion disk. However  it depends on the several
 simplifications which are subject to be verified. For example, we use the
 relation of the field components suggested by  Blandford
 which requires a justification if it can be used  in
the relativistic formulation especially  for a rapidly rotating
black hole.  Also the identification of the angular velocity of the
field line as the Keplerian angular velocity in this model also
needs a valid justification\cite{hkl}.

\vskip 0.5cm
This work was supported by Hanyang University, Korea made in the program year of 2001.

%%%%%%%%%%%%%%%%%%%%%%%%%%%%%%%%%%%%%%%%%%%%%%%%%%%%%%%%%%%%%%%%
%%%%%%%%%%%%%%%%%%%%       table      %%%%%%%%%%%%%%%%%%%%%%%%%%
%%%%%%%%%%%%%%%%%%%%%%%%%%%%%%%%%%%%%%%%%%%%%%%%%%%%%%%%%%%%%%%%
\newpage

\begin{table}
\begin{tabular}{|c|c|c|c|c|} \hline
GRB & z & $d_{L}[cm]$ & $E_{iso}[erg]$ & $\tilde{a}_{l}$ \\ \hline
970228 & 0.695 & 1.403e+28 & 1.42e+52 & 0.14 \\ \hline
970508 & 0.835 & 1.757e+28 & 5.46e+51 & 0.09 \\ \hline
970828 & 0.958 & 2.082e+28 & 2.20e+53 & 0.31 \\ \hline
971214 & 3.418 & 9.877e+28 & 2.11e+53 & 0.31 \\ \hline
980613 & 1.0969 & 2.459e+28 & 5.67e+51 & 0.09 \\ \hline
980703 & 0.9662 & 2.103e+28 & 6.01e+52 & 0.21 \\ \hline
990123 & 1.6004 & 3.925e+28 & 1.44e+54 & 0.45 \\ \hline
990506 & 1.30 & 3.037e+28 & 8.54e+53 & 0.42 \\ \hline
990510 & 1.619 & 3.982e+28 & 1.76e+53 & 0.30 \\ \hline
990705 & 0.84 & 1.770e+28 & 2.70e+53 & 0.33 \\ \hline
990712 & 0.433 & 7.927e+27 & 5.27e+51 & 0.09 \\ \hline
991208 & 0.707 & 1.429e+28 & 1.47e+53 & 0.28 \\ \hline
991216 & 1.02 & 2.250e+28 & 5.35e+53 & 0.39 \\ \hline
000131 & 4.500 & 1.369e+29 & 1.16e+54 & 0.44 \\ \hline
000301C & 2.034 & 5.269e+28 & 4.64e+52 & 0.19 \\ \hline
000418 & 1.118 & 2.523e+28 & 8.29e+52 & 0.24 \\ \hline
000926 & 2.037 & 5.280e+28 & 2.97e+53 & 0.34 \\ \hline
\end{tabular}
\caption{The isotropic energies of GRB[10] and the lower limits of angular momentum parameter, $\tilde{a}_{l}$.}
\end{table}

\newpage

\begin{table}
\begin{tabular}{|c|c|c|c|c|c|c|c|} \hline
\multicolumn{2}{|c|}{GRB} & 990123 & 990506 &990510 & 991208 & 991216 &000131 \\ \hline
\multicolumn{2}{|c|}{red shift, z} & 1.6004 & 1.3 & 1.619 & 0.707 & 1.02 & 4.5 \\ \hline
\multicolumn{2}{|c|}{$E_{iso}[M_{\odot}]$} & 0.7167 & 0.395 & 0.0678 & 0.0644 & 0.2417 & 0.572 \\ \hline
\multicolumn{2}{|c|}{$T_{90}[s]$} & 24.34 & 57.10 & 25.80 & 39.84 & 7.51 & 9.09 \\ \hline
& $B_{H}(0)[10^{15}gauss]$ & 5.19 & 2.56 & 1.49 & 1.17 & 5.44 & 7.69 \\ \cline{2-8}
$\tilde{a}(0)=1$ & $M_{D}(0)[M_{\odot}]$ & 3.28 & 1.54 & 0.16 & 0.15 & 0.82 & 2.47 \\ \cline{2-8}
& $M^{f}_{BH}[M_{\odot}]$ & 8.71 & 7.71 & 7.04 & 7.03 & 7.33 & 8.24 \\ \cline{2-8}
& $\tilde{a}^{f}$ & 0.80 & 0.87 & 0.98 & 0.98 & 0.92 & 0.83 \\ \hline
& $B_{H}(0)[10^{15}gauss]$ & 5.96 & 3.21 & 2.21 & 1.73 & 7.29 & 9.13 \\ \cline{2-8}
$\tilde{a}(0)=0.8$ & $M_{D}(0) [M_{\odot}]$ & 4.36 & 2.34 & 0.39 & 0.37 & 1.41 & 3.44 \\ \cline{2-8}
& $M^{f}_{BH}[M_{\odot}]$ & 9.75 & 8.46 & 7.24 & 7.23 & 7.87 & 9.16 \\ \cline{2-8}
& $\tilde{a}^{f}$ & 0.71 & 0.74 & 0.79 & 0.79 & 0.76 & 0.72 \\ \hline
&$B_{H}(0)[10^{15}gauss]$ & & & 7.31 & 5.68 & & \\ \cline{2-8}
$\tilde{a}(0)=0.3$ & $M_{D}(0) [M_{\odot}]$ & & & 0.59 & 0.56 & & \\ \cline{2-8}
& $M^{f}_{BH}[M_{\odot}]$ & & & 7.42 & 7.40 & & \\ \cline{2-8}
& $\tilde{a}^{f}$ & & & 0.22 & 0.23 & & \\ \hline
\end{tabular}
\caption{The initial disk mass and the magnetic field of black hole for GRBs
for which $T_{90}$s are well defined.
$M_{BH}^{f}$ and $\tilde{a}^{f}$ are the final black hole mass and
the angular momentum parameter respectively.
The energetic GRBs can not accomodate the black hole with $\tilde{a}(0)=0.3.$}
\end{table}

%%%%%%%%%%%%%%%%%%%%%%%%%%%%%%%%%%%%%%%%%%%%%%%%%%%%%%%%%%%%%%%%
%%%%%%%%     Figures    %%%%%%%%%%%%%%%%%%%%%%%%%%%%%%%%%%%%%%%%
%%%%%%%%%%%%%%%%%%%%%%%%%%%%%%%%%%%%%%%%%%%%%%%%%%%%%%%%%%%%%%%%

\newpage
\begin{figure}[tbp]
  \includegraphics[height=12cm]{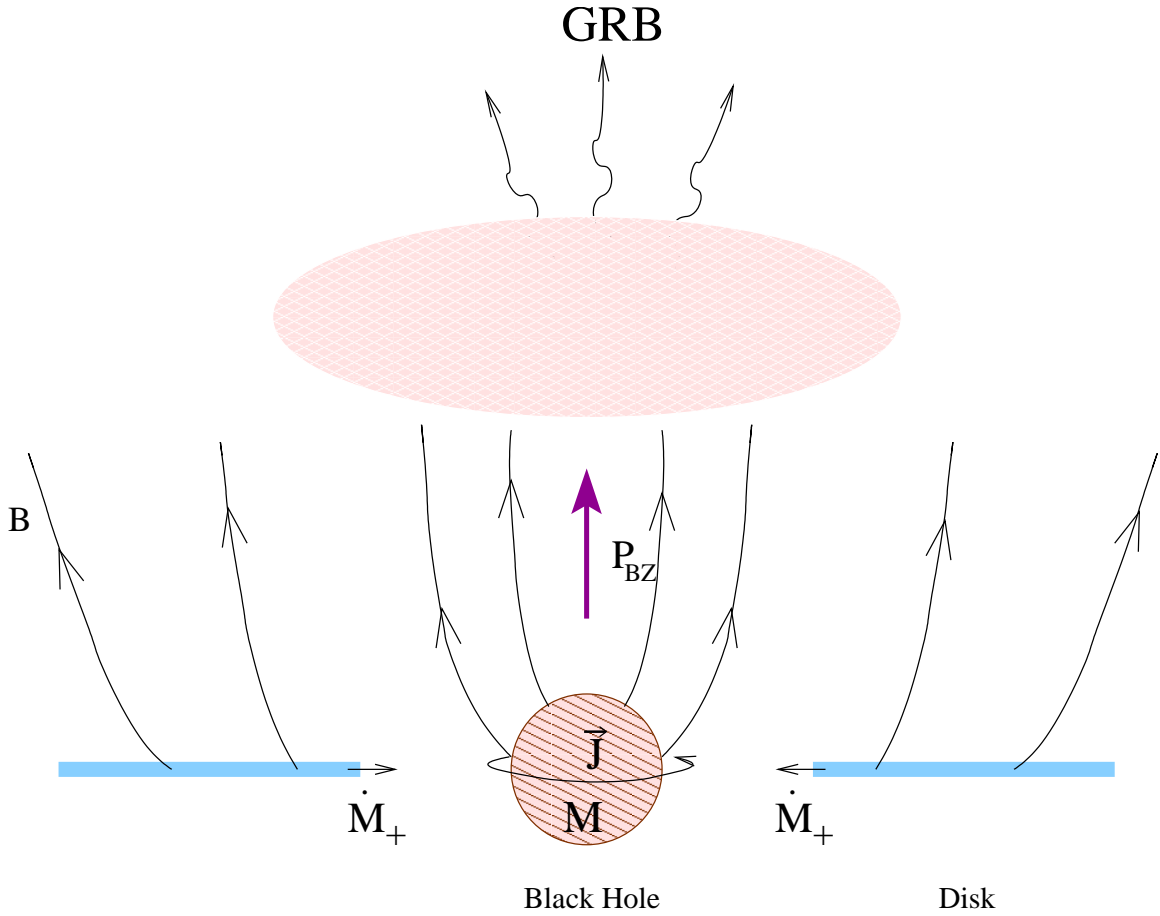}
  \caption{Skematic representation of the black hole - accretion disk system powering GRB(only upper half plane is shown).
The rotational energy extracted out by poynting flux $P_{BZ}$
is supposed to power GRB.
The evolution of the black hole with the mass M and the angular momentum J
is determined by $P_{BZ}$ and the accretion rate $\dot{M}_{+}$.}
\end{figure}

\newpage
\begin{figure}[tbp]
  \includegraphics[height=11cm]{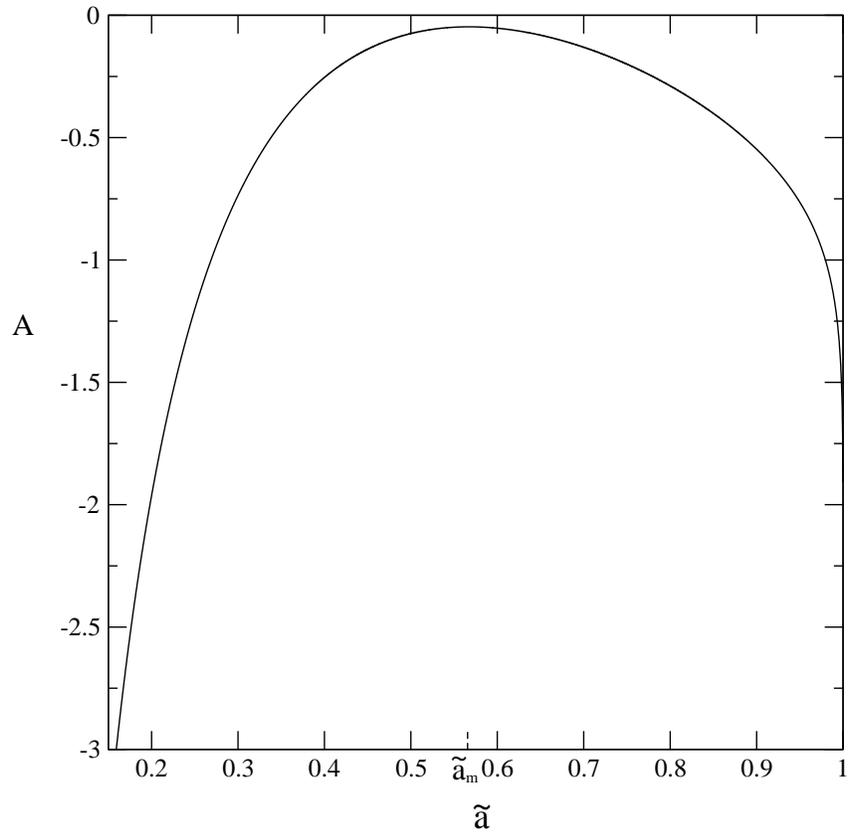}
  \caption{Value of the sign factor A, which has a
maximum at $\tilde{a}_{m}=0.57$.}
\end{figure}

\newpage
\begin{figure}[tbp]
  \includegraphics[height=12cm]{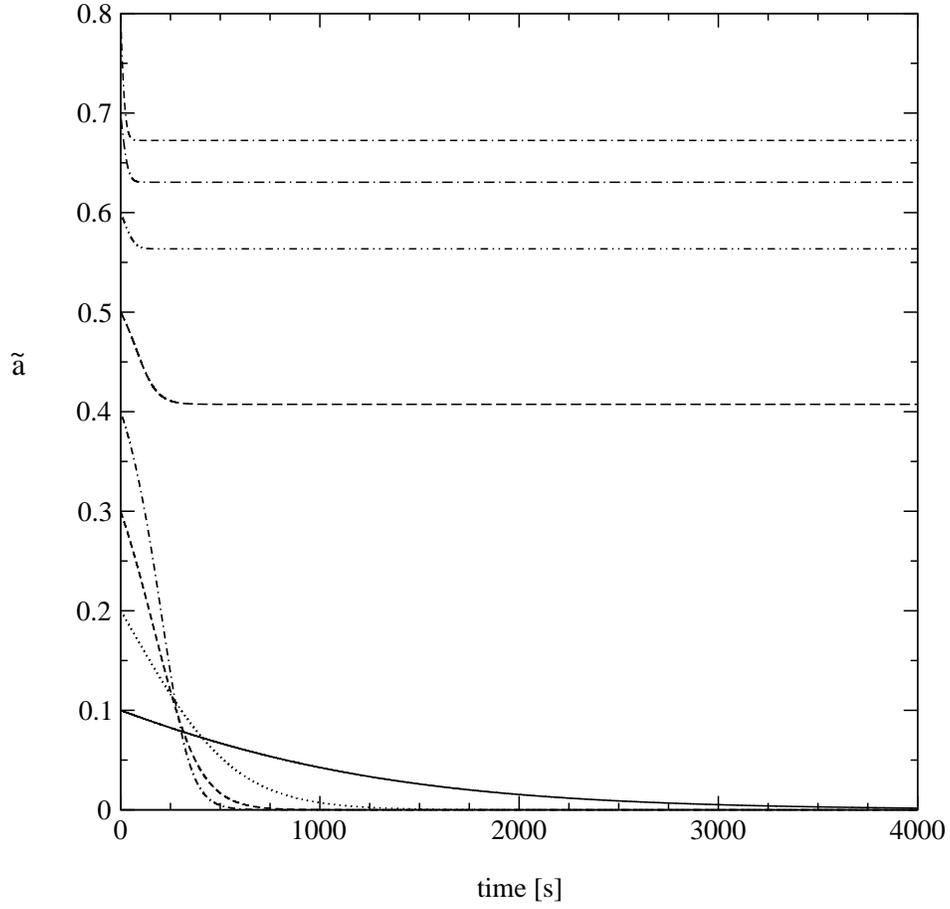}
  \caption{The evolutions of the angular momentum parameter in time with $M(0) = 7M_{\odot}$ and $M_{D}(0) = 3M_{\odot}$.}
\end{figure}

\newpage
\begin{figure}[tbp]
  \includegraphics[height=11cm]{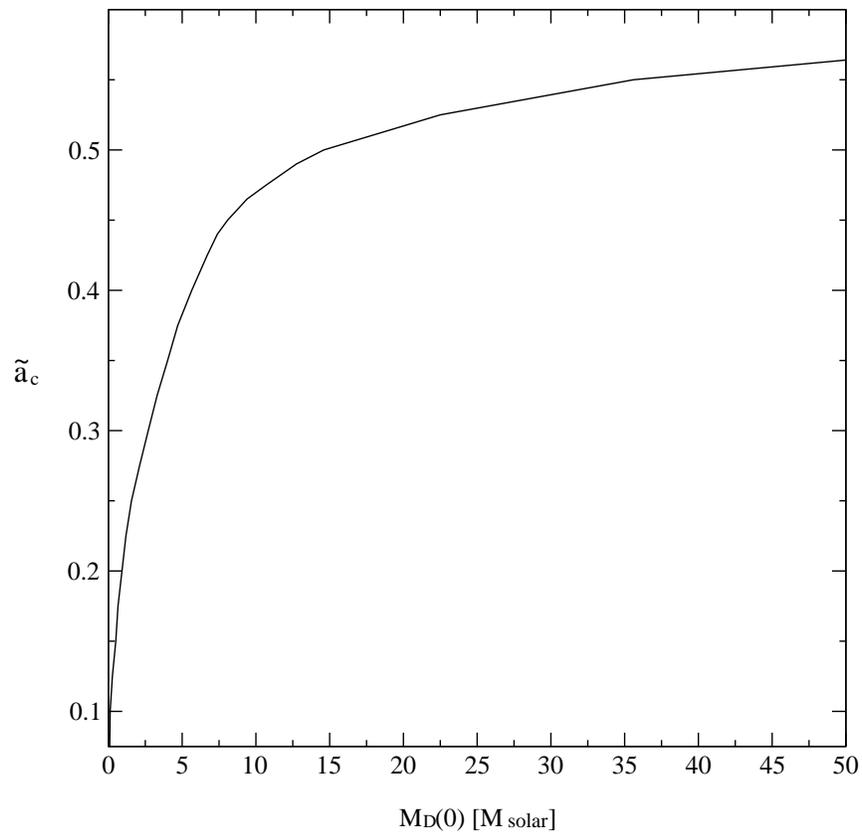}
  \caption{$\tilde{a}_{c}$ for a given mass of accretion disk in the unit of solar mass $M_{solar}$.}
\end{figure}

\newpage
\begin{figure}[tbp]
    \includegraphics[height=12cm]{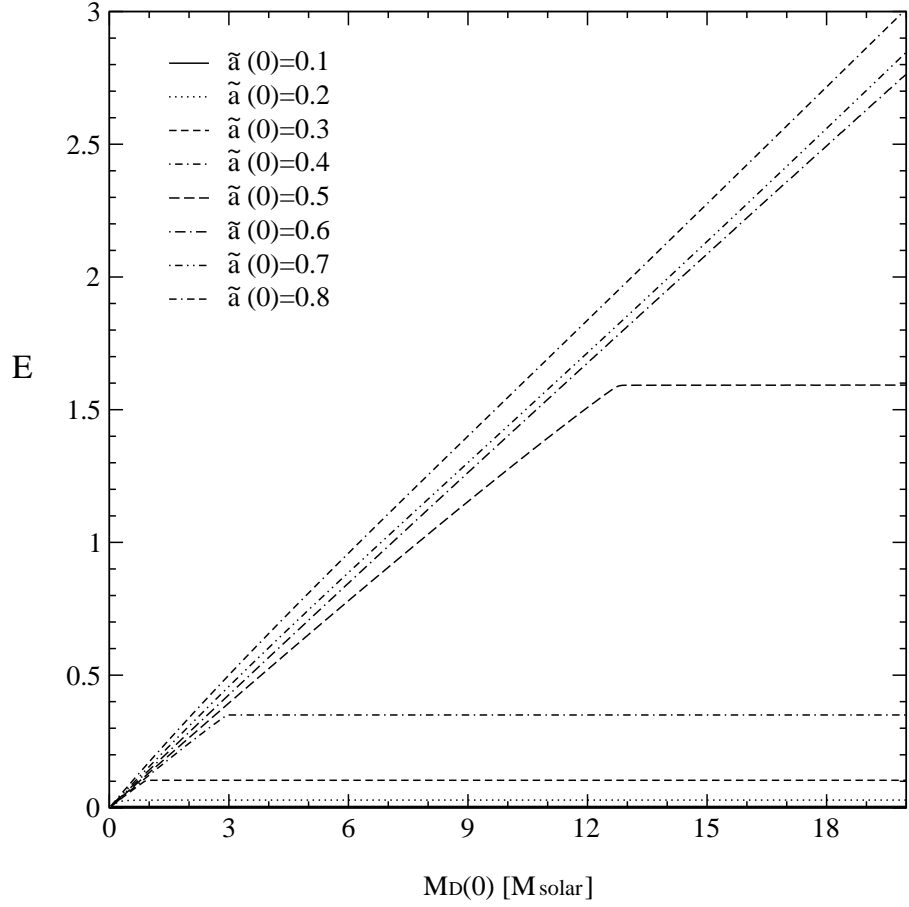}
    \caption{The energy vs. the initial disk mass($M_{D}(0)$)
for various initial angular momentum parameters($\tilde{a}(0)$).}
\end{figure}

\newpage
\begin{figure}[tbp]
  \includegraphics[height=11cm]{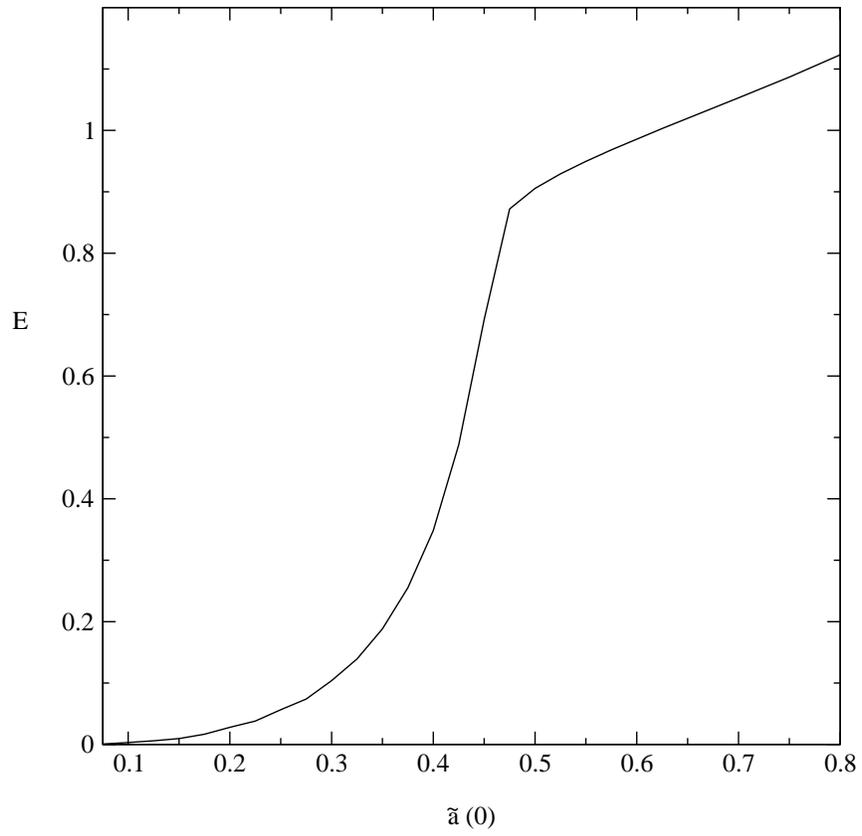}
  \caption{The upper limits of the energy out of the system
as a funtion of $\tilde{a}(0)$.}
\end{figure}

\newpage
\begin{figure}[tbp]
  \includegraphics[height=12cm]{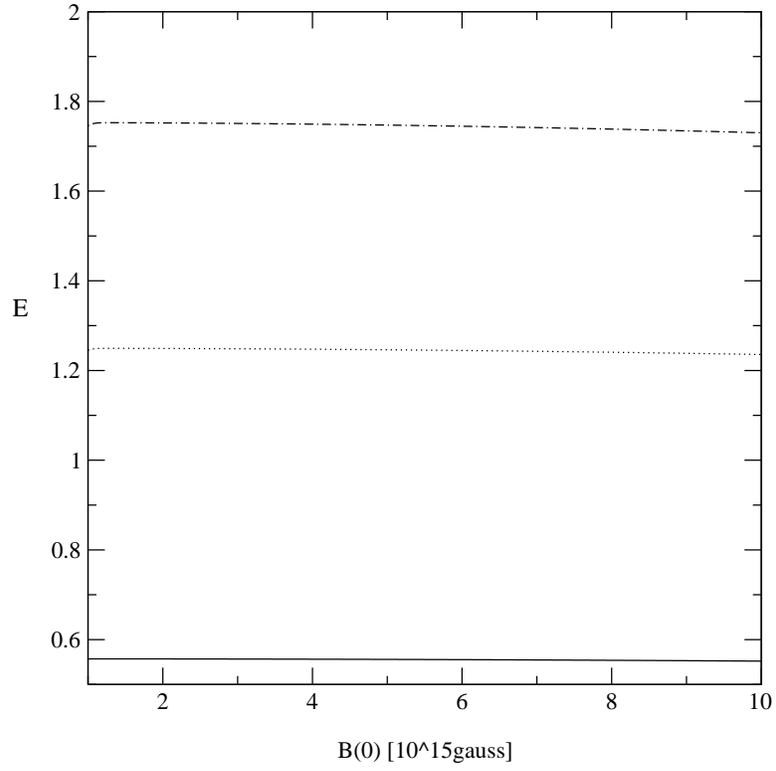}
  \caption{The effect of the magnetic field on the total energy out of
the system for  $M_{D}(0) = 3M_{\odot}$(solid line) and
$M_{D}(0) = 7M_{\odot}$(dotted line) and $M_{D}(0) = 10M_{\odot}$
(solid - dotted line).}
\end{figure}

\end{document}